\documentclass{mem}
\usepackage{natbib}\usepackage{txfonts}\usepackage{balance}
\usepackage{graphicx}
\usepackage[a4paper,breaklinks,dvipdfm]{hyperref}
\idline{75}{282}
\begin{document}
\def\teff{$T\rm_{eff }$}
\def\kms{$\mathrm {km s}^{-1}$}

\title{
THESEUS and the high redshift universe
}

   \subtitle{}

\author{
N.\,R.\,Tanvir\inst{1} 
          }

\institute{
Department of Physics and Astronomy, University of Leicester,
University Road, Leicester, LE1 7RH. United Kingdom
\email{nrt3@le.ac.uk}
}

\authorrunning{Tanvir}

\titlerunning{THESEUS and the high redshift universe}

\abstract{
Long-duration gamma-ray bursts (long-GRBs) can be detected throughout cosmic history and provide
several unique insights into star-formation and galaxy evolution back to the era of
reionization.  They can be used to map star formation, identify galaxies across the luminosity function,
determine detailed abundances even for the faintest of galaxies, quantify the escape fraction of ionizing 
radiation and track the progress of reionization.
Fully exploiting these techniques requires a significant increase in the number
of long-GRBs identified and characterised at $z\gtrsim6$, which can be achieved through a
discovery mission with the capabilities of {\em THESEUS}, in combination with the powerful follow-up facilities that will be
available in the 2030s.
\keywords{
Galaxies: abundances -- Cosmology: observations }
}
\maketitle{}

\section{Introduction}

Understanding the evolution of the early generations of stars and galaxies in the universe, and the 
accompanying reionization of the intergalactic medium, 
are primary objectives of contemporary astrophysics.  In particular, whether extreme ultra-violet radiation
from those early stars was the predominant driver of
reionization is a crucial question, since if not then some other substantial source of ionizing radiation must be found
\citep[e.g.][]{Robertson2010}.
Determination of the electron scattering optical depth from microwave background observations
indicate a peak era of reionization around $z\sim7$--10 \citep{Planck2016}.
Considerable progress has been made in recent years in unveiling the galaxy
populations at these redshifts, particularly thanks to the various
deep field campaigns undertaken with the {\em Hubble Space Telescope} \citep[e.g.][]{Koekemoer2013}, most recently the
Frontier Fields initiative employing gravitational lensing to probe to fainter levels than would
otherwise be possible \citep[e.g.][]{Ishigaki2015}.  
This has suggested that a major proportion of star formation is occurring 
in very faint galaxies \citep[e.g.][]{Atek2015}, for which direct 
constraints on their number and properties are very limited.
\begin{figure*}[t!]
\begin{minipage}{70mm}
\includegraphics[clip=true,angle=90,width=66mm]{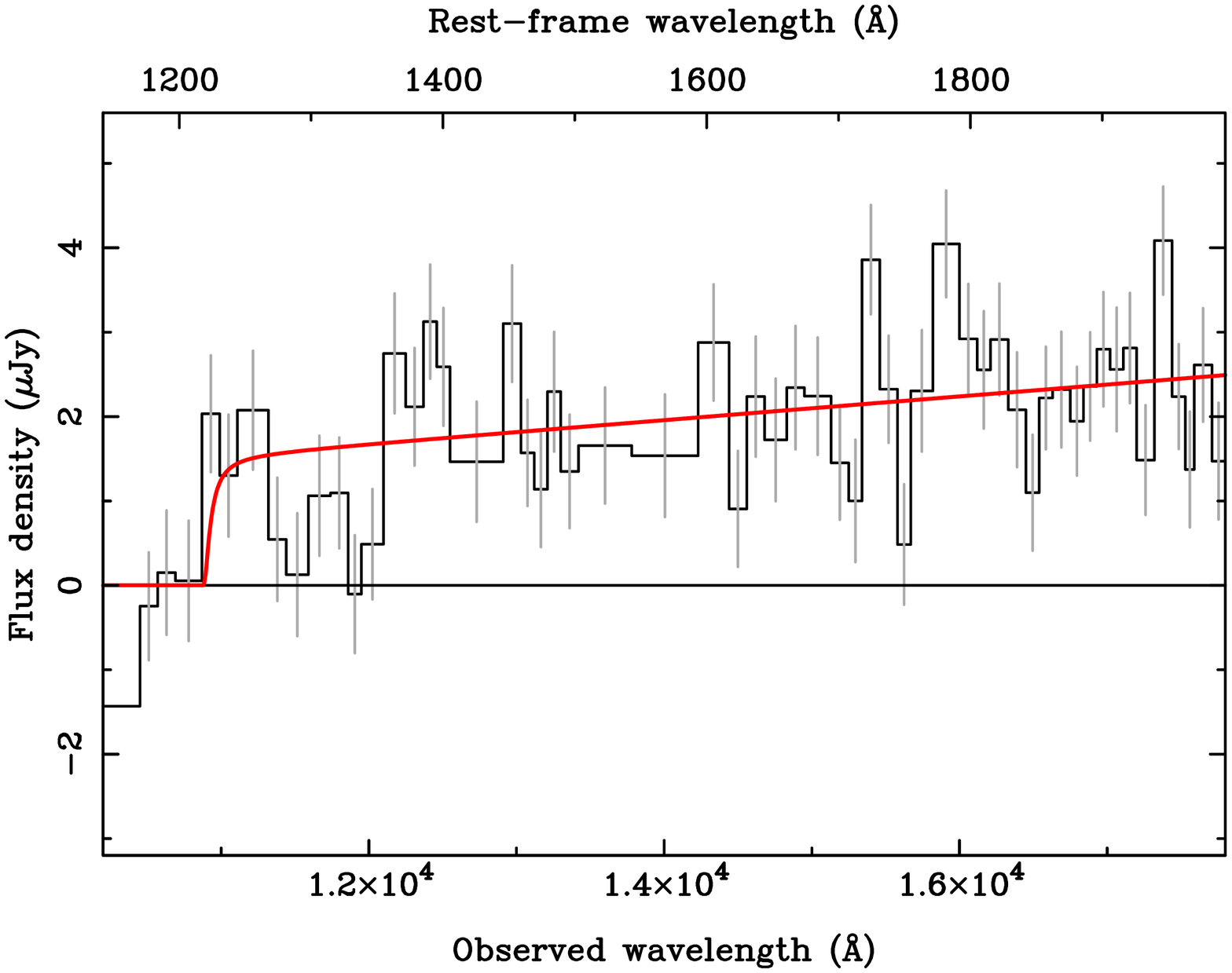}
\end{minipage}
\begin{minipage}{67mm}
\vspace{5mm}
\includegraphics[clip=true,angle=270,width=64mm]{120923A_sim.ps}
\end{minipage}
\caption{\footnotesize
Left - the observed spectrum of GRB\,120923A \citep{Tanvir2017} which
was obtained from a 2\,hr VLT X-shooter spectrum.  This afterglow was particularly
faint and challenging for current technology, and only allowed the redshift, $z\sim7.8$, to be determined
from the Ly-$\alpha$ break.  Right - a simulated ELT/HARMONI spectrum
of the same afterglow, illustrating the huge improvement in signal-to-noise expected,
and consequent detection of metal lines and precise determination of the HI column
from the fit to the Ly-$\alpha$ damping wing.
}
\label{120923A}
\end{figure*}

Several {\em Swift}-discovered long-duration gamma-ray bursts (long-GRBs) 
have been found approaching and within the era of reionization \citep[e.g.][]{Tanvir2009,Salvaterra2009,Cucchiara2011,Tanvir2017}.
As outlined in this brief contribution, these high redshift long-GRBs have already produced unique insights into 
high-$z$ star formation, and have paved the way for  the key high-$z$ science theme of 
the {\em THESEUS} mission.

\section{The astronomical landscape of the late 2020s}

Gamma-ray bursts are quintessential multi-wavelength, and indeed multi-messenger,
phenomena, and so the scientific return obtained from GRB missions is enhanced
greatly by the facilities available for complementary observations and follow-up.
By the time {\em THESEUS} is operational, if selected for an M5 launch, the landscape of 
astrophysical hardware is likely to be significantly different from today.
Supplementing the current generations of 8\,m class optical telescopes, 
ground-based radio, submm, and gravitational wave detectors, 
together with the X-ray, gamma-ray and optical observatories in space,
we expect a new generation of 30\,m class ground-based optical/IR telescopes,
the Square Kilometre Array, potentially third-generation gravitational wave
detectors, the Large Synoptic Survey Telescope, and {\em ATHENA} in space.

The {\em THESEUS} mission will provide the essential link to exploit the synergies 
between these facilities for transient science generally, and in the exploration of
the early universe in particular.

\section{The role of GRBs}

Long-duration GRBs are found over a large span of cosmic history; they are 
born in massive star core-collapse, and lie at the star-forming hearts of galaxies.  
Thus they provide a range of unique probes high redshift galaxy evolution, which
will be exploited by the {\em THESEUS} mission together with follow-up observations.

\subsection{Evolution of the global star formation rate density}
Since long-GRBs are core-collapse phenomena, they trace massive star formation \citep[e.g.][]{Blain2000}.
Thus the observed GRB redshift distribution, providing the sample is redshift complete, 
can in principle be inverted to estimate the global star formation evolution without 
regard to whether the host galaxies are detected or not.
Early attempts already showed that the long-GRB rate was surprisingly high at $z>4$,
given the fairly rapid decrease in star formation being found by galaxy surveys \citep[e.g.][]{Kistler2009}.
In practice, it is known that long-GRBs are preferentially created in lower metallicity
environments, and suitable accounting for this effect, combined with a realisation that a
greater proportion of high-z star formation is likely happening in very faint galaxies,
has brought estimates into better agreement \citep{Perley2016}.
However, this remains a critical question for reionization.

\subsection{Locating star forming galaxies}
By virtue of localising GRB afterglows and determining their redshifts, we can sample
faint galaxy populations independently of their luminosities {\citep[e.g.][]{McGuire2016}. 
This is in contrast to conventional galaxy
surveys, that of course depend on detecting the galaxies in some band(s), and in the large
majority of cases rely on photometric redshifts.
This is a particular issue at $z\gtrsim6$, where the galaxy luminosity function becomes increasingly
steep and faint-end dominated, and corrections for these missed galaxies difficult 
and uncertain \citep[e.g.][]{Bouwens2017}.
By comparing the number of GRBs in hosts above some given detection threshold to the
number below, one can directly estimate the correction factor for the proportion of star formation
missed in conventional galaxy surveys
\citep{Tanvir2012,Trenti2012,Basa2012}.
\begin{figure*}[t!]
\centerline{\includegraphics[clip=true,angle=270,width=110mm]{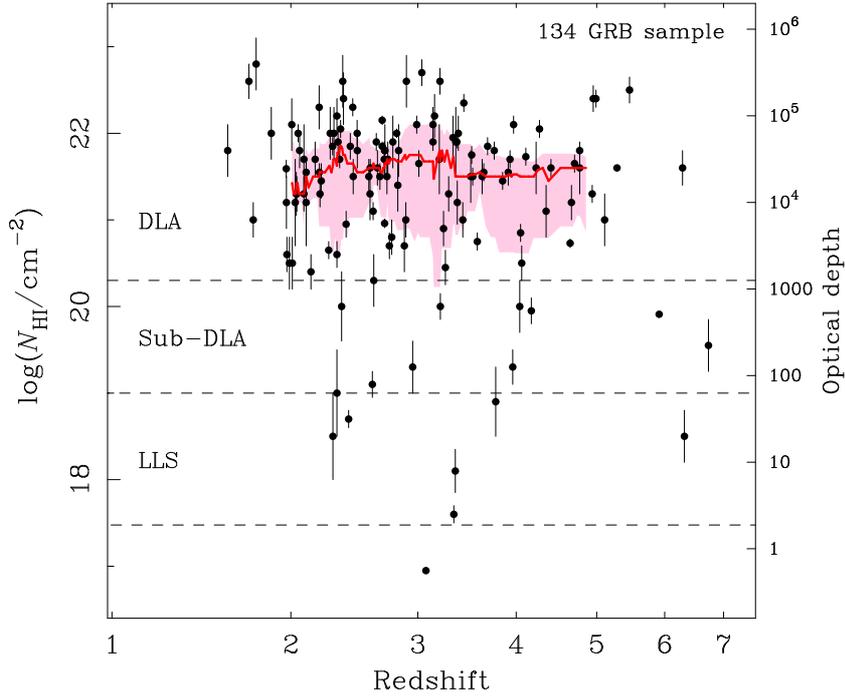}}
\caption{\footnotesize
Neutral hydrogen column density measured from fits to the Lyman-$\alpha$
absorption lines in GRB afterglow spectra.  On the right-hand axis is shown the
corresponding optical depth to H-ionizing extreme ultra-violet radiation. In nearly all cases the sight-lines
are essentially opaque, implying a very low escape fraction.
The running median and interquartile range of 20 events (red line and pink band) show little
evidence for evolution between $2\lesssim z\lesssim5$. (Figure based on data
presented in Tanvir et al. submitted.)}
\label{nhz}
\end{figure*}

\subsection{Cosmic chemical evolution}
GRB afterglows provide bright back-lights against which numerous absorption features
created by intervening gas clouds in the host can often be seen.  These  provide 
not only gross metallicities, but detailed abundance patterns from which the enrichment by
prior generations of stars can be inferred.
Once again, this is independent of the host luminosity, unlike crude emission line
diagnostics, and can be applied even at high redshifts 
\citep{Vreeswijk2004,Thoene2013,Sparre2014,Hartoog2015}.
The complementarity of {\em THESEUS} GRB discoveries, and optical/nIR followup with
30-m class telescopes promises a major step forward, which is illustrated by
the comparison of the VLT spectrum of GRB\,120923A and simulated ELT spectrum
of the same afterglow in Figure~\ref{120923A}.

\subsection{Ionizing radiation escape fraction}
Direct observation of escaping Lyman continuum radiation from distant galaxies is challenging
at $z\sim2$--4 \citep[e.g.][]{Japelj2017}, and essentially impossible at higher redshift due to strong IGM absorption.
Afterglow spectra of GRBs frequently exhibit strong absorption due to hydrogen Lyman-$\alpha$,
which allows calculation of neutral hydrogen column density, and hence the opacity to
ionizing radiation with $\lambda<912$\,\AA\ \citep{Chen2007,Fynbo2009}.
As can be seen from Fig.~\ref{nhz}, over a wide range of redshift the large majority of GRB sight-lines
are essentially opaque, and thus the bulk of ionizing radiation from the progenitor stars
would not escape the host galaxies.
Assuming these sight-lines are representative of the sight-lines to massive stars more generally, one
can thus infer an average escape fraction of  ionizing radiation, which, in a new study by Tanvir et al. (MNRAS submitted)
is found to have a 98\% upper limit of 1.5\%.
This is potentially a problem for the hypothesis that reionization was brought about by UV from 
massive star, since that seems to require escape fractions of at least 10--20\%.
Although this GRB sample is largely in the range $2\lesssim z \lesssim5$, there is little
evidence of variation with redshift.
From the  {\rm THESEUS} mission we expect to greatly increase the sample of $z>5$ GRBs
with precise $N_{\rm HI}$ measures, thus providing a strong test of whether sufficient stellar ionizaing
radiation can escape from the locations of massive stars to drive reionization.

\subsection{Topology of reionization}
Just as the neutral gas in the host interstellar medium on the lines-of-sight to GRBs can be
inferred from the Lyman-$\alpha$ absorption line, so any neutral gas in the intergalactic
medium (IGM) proximate to the host also contributes to the absorption. 
The shape of the damping wing in each case differs slightly, since the IGM absorption
is an integrated effect of gas over a path length through the expanding universe.
Thus in principle the two columns can be decomposed and hence the neutral fraction 
of the IGM at that location estimated.  
In practice, the method is hard for low signal-to-noise spectra, and may be complicated
by `proximity effects' such as inflows or outflows and local ionized regions  
around the host \citep{McQuinn2008}, although these
effects are likely much less significant than in the case of bright quasars.
This approach has been attempted in a couple of cases to date \citep{Totani2006,Hartoog2015},
with results consistent with a low neutral fraction at $z\sim6$.
By obtaining similar results for a larger sample of sight-lines in the {\em THESEUS} era
we will be able to investigate both the overall timeline, but also the variation
from place to place (and hence the topology) of reionization.

\subsection{Population III stars}
Inefficient cooling of metal-free gas tends to produce stars with a top-heavy initial mass function
\citep[e.g.][]{Stacy2016}, and if some of these very massive pop III stars end their lives with 
high specific angular momentum they may produce energetic collapsars.
If the jets they produce are sufficiently  long-lived
 then they may produce a distinct class of pop-III GRBs \citep{Meszaros2010,Yoon2015}.
Even if not detectable directly, the chemical signatures of pop III enrichment
may be witnessed in spectroscopy of high-redshift GRBs \citep{Ma2015}.

\section{Conclusions}

The study of distant galaxy populations, and their role in the reionization of the universe, 
have been the subject of major efforts, and are a primary science driver for {\em JWST}.
Despite this, some key questions will continue to be very hard to answer, in particular the star formation occurring
in faint galaxies, the build up of heavy elements and the escape fraction of ionizing radiation.
Long-duration gamma-ray bursts provide unique routes to investigate these issues, which will be
fully exploited using the large samples of high-$z$ GRBs found by {\em THESEUS} together with follow-up by
next generation facilities.


\bibliographystyle{aa}

\end{document}